\documentclass[mathleft
]{an}
\usepackage{graphicx}
\usepackage{times}
\begin{document}

\Pagespan{1}{}
\Yearpublication{2011}%
\Yearsubmission{2011}%
\Month{11}%
\Volume{999}%
\Issue{88}%

\title{The short-period low-mass binary system CC Com revisited}

\author{O. K\"ose\inst{1}
\and  B. Kalomeni\inst{1,2} \and  V. Keskin\inst{1} \and  B.Ula{\c s}\inst{1} \and  K. Yakut\inst{1}
}
\titlerunning{Short period close binary system CC Com}
\authorrunning{K\"ose et al.}
\institute{
Department of Astronomy and Space Sciences, University of Ege, 35100, {\.I}zmir, Turkey
\and
Department of Physics, \.Izmir Institute of Technology, 35430, {\.I}zmir, Turkey }

\received{\today}
\accepted{\today}
\publonline{later}

\keywords{stars: binaries: close; stars: individual (CC Com); stars: low-mass; stars: evolution}

\abstract{%
In this study we determined precise orbital and physical parameters of the very short period low-mass contact binary system CC Com.
The parameters are obtained by analysis of the new CCD data with the archival spectroscopic data. The physical parameters of the
components derived as $M_\textrm{c}$ = 0.717(14) $M_{\odot}$, $M_\textrm{h}$ = 0.378(8) $M_{\odot}$, $R_\textrm{c}$ = 0.708(12) $R_{\odot}$,
$R_\textrm{h}$ = 0.530(10) $R_{\odot}$, $L_\textrm{c}$ = 0.138(12) $L_{\odot}$, $L_\textrm{h}$ = 0.085(7) $L_{\odot}$, and the distance of the system is estimated as 64(4) pc.
The times of minima obtained in this study and with those published before enable us to calculate the
mass transfer rate between the components which is  $1.6\times10^{-8}$ M$_{\odot}$yr$^{-1}$.
Finally, we discuss the possible evolutionary scenario of CC Com.}

\maketitle

\section{Introduction}

One of the crucial parameter that determines the evolutionary stages of a binary is its orbital parameter.
Because of their unusual behaviours short period systems like CC Com, GSC 1387-0475 (Yang et al. 2009) and V523 Cas (K\"{o}se et al. 2009) are important in evolutionary studies.

CC Com was discovered by Hoffmeister (1964) and has been intensively studied photometrically
by Rucinski (1976),  Breinhorst \& Hoffmann (1982), Bradstreet (1985), Zhou (1988),
Linnell \& Olson (1989) and Zola et al. (2010). Over the last four decades an intense spectroscopic studies were made by
Rucinski et al. (1977), McLean \& Hilditch (1983), and Pribulla et al. (2007). In this studies spectroscopic mass ratio was
found as 0.52(3), 0.47(4), and 0.53(1), respectively. Photometric mass ratio, on the other hand, was given as 0.59 by
Zhou (1988) and 0.51 by Linnell \& Olson (1989). The physical parameters of the components have not been measured by
simultaneous analysis of spectroscopic and photometric data.

Orbital period study of weakly contact binary CC Com has been the subject of many papers. These studies indicate a systematic period decrease.
In the literature the period variation ($dP/dt$) of CC Com was given as  $-4.4\times10^{-8}$ d~yr$^{-1}$, $-4.0\times10^{-8}$ d~yr$^{-1}$,
and $-2.0\times10^{-8}$ d~yr$^{-1}$ by Qian (2001a), Yang \& Liu (2003), and Yang et al. (2009), respectively.
Yang et al. (2009) presented cyclic variations superimposed on a parabolic period variation and discussed this feature as an indication of a third body or stellar activity.

Contact model solution has been suggested by Rucinski (1976) and related UBV parameters have been given in that study.
Breinhorst \& Hoffmann (1982) studied the effects that can cause variations in minima depths assuming no period change.
The filling factor variation has been discussed by Linnell \& Olson (1989) based on their $u$, $y$ and $I$ photometric data analysis.
Recently, Zola et al. (2010) gave the photometric elements of the system by using Wilson-Devinney code.

In this paper, we present photometric analysis and orbital period study of the short-period
eclipsing binary system CC Com. First, the new observations with new linear ephemeris are presented.
Then the mass transfer rate between the components is estimated by period study.
Next, the light curve of the system is modeled by comparing the results with the results of previous works.
Finally, obtained results are presented with a discussion of possible evolutionary stages of the binary.

\section{New Observations}

\begin{figure}
\includegraphics[height=90mm]{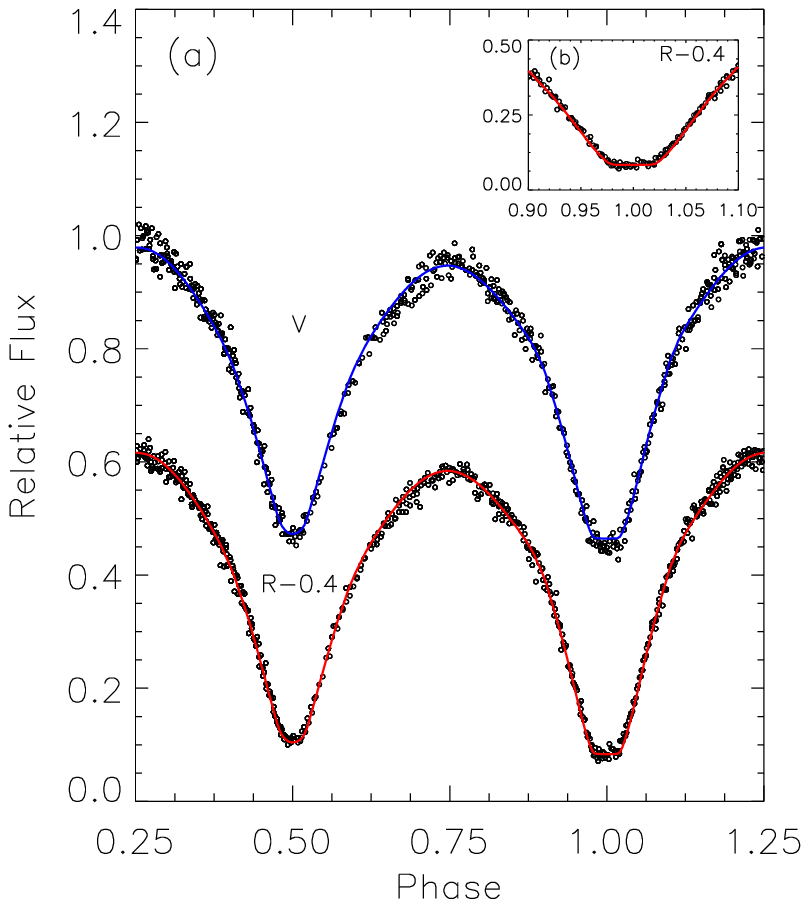}\\
\includegraphics[height=45mm]{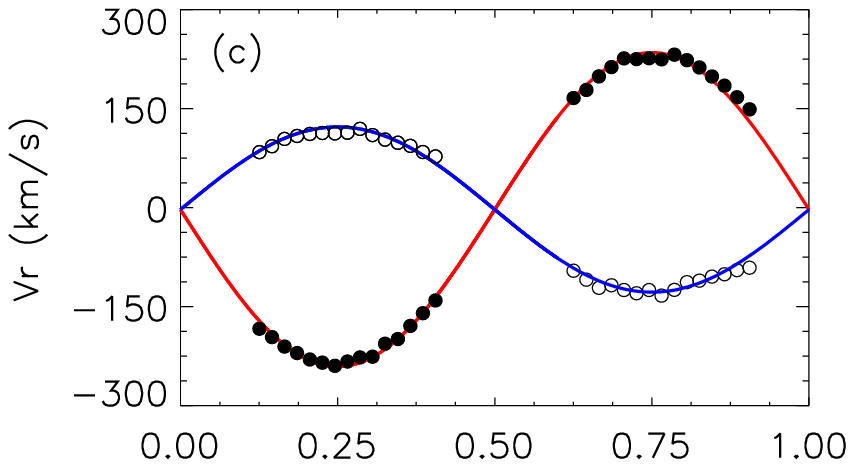}
\caption{(a) The observed and the computed (solid line) light curves of CC Com.
The light curve in R band is moved by a value of -0.4 in intensity, (b) minimum light is zoomed between phases 0.90 and 1.10 for a good visibility,
(c) Radial velocities of CC Com. The data obtained from Pribulla et al. (2007). The computed lines are estimated from simultaneous solution.}\label{fig1}
\end{figure}

We observed CC Com in the Bessel V and R filters in 2007 on six nights by using
40-cm telescope with an Apogee U47 CCD at T\"UB\.ITAK National Observatory (TUG) and one night in 2011 at Ege University Observatory
with the 40-cm telescope equipped with the Apogee CCD camera.
GSC 01986 01673  and GSC 01444 00106 are chosen as comparison and check stars, respectively.
The integration was 30~s in V and 25~s in R bands.
IRAF (DIGIPHOT/ APPHOT) packages are used to reduce the CCD data.
The errors are 0.011 mag in V and 0.008 mag in R passbands.
The light-curve of the binary which shows a total eclipse is shown in Fig.~\ref{fig1}.
In Table~\ref{tab:CCCom:mintimes} newly obtained six times of minima are listed with those published in the literature.

Period variation study of CC Com performed by using a total of 83 collected times of minima light obtained by photometric/CCD
observations. The new linear ephemeris derived in this study is
\begin{eqnarray}
\label{cccomEq1}
\rm{HJD\,MinI}  = 24\,54\,151.6060(2) +  0.22068516(6)\times E.
\end{eqnarray}

\section{Eclipse timings and period study}
CC Com is a contact binary system, in which continuous mass-transfer between the components is expected.
A parabolic variation can be seen in the O-C diagram due to the adequately high mass transfer rate
that can be derived by period analysis. Period variation of CC Com has been discussed in some papers (Qian 2001a, Yang \& Liu 2003, and Yang et al. 2009).
Qian (2001a) based on 35 photometric/CCD minima times studied the parabola like variation and gave the quadratic term ($Q$) as $-1.323(3)\times10^{-11}$.
Yang \& Liu (2003) analyzed 322 times of minima light including the visual data assuming parabolic variation the
$Q$ value was obtained to be $-1.2(4)\times10^{-11}$ (\textit{ibid}).
Recently, Yang et al. (2009) by using visual and photometric data obtained a sine-like variation
superimposed on a parabolic variation ($Q=-0.59(5)\times10^{-11}$). The period of sine-like
variation was calculated as 23.6 yrs.
This effect has been interpreted as an indication of either cyclic stellar activity or a third body in the system.

The scattering in visual data is about 0.03 days. This makes it difficult to study any low amplitude variation.
Hence, visual data points are excluded in period analysis. Times of mid-eclipses with those
obtained in this study are given in Table~\ref{tab:CCCom:mintimes}.

\begin{table}
\begin{center}
\caption{The times of minimum light of CC Com. The data obtained before 2001 was given in Qian (2001a).}\label{tab:CCCom:mintimes}
\begin{tabular}{llll}
\hline
HJD* Min &      Ref &   HJD Min &      Ref \\
\hline
52002.3484	&	1	&	53823.7849	&	11	\\
52002.4592	&	1	&	53824.7758	&	11	\\
52039.4238	&	1	&	53847.3921	&	7	\\
52648.9580	&	2	&	53850.3695	&	12	\\
52721.3429	&	3	&	54175.4396	&	10	\\
52800.0165	&	4	&	54175.5505	&	10	\\
53068.5921	&	5	&	54198.3907	&	13	\\
53093.4195	&	5	&	54202.3634	&	10	\\
53093.5298	&	5	&	54203.3558	&	14	\\
53106.4436	&	6	&	54204.3531	&	10	\\
53116.4819	&	6	&	54206.3358	&	10	\\
53122.3280	&	5	&	54209.4245	&	10	\\
53446.4060	&	7	&	54209.5347	&	10	\\
53460.1993	&	4	&	54213.3979	&	13	\\
53460.3101	&	4	&	54593.4190	&	14	\\
53460.5312	&	6	&	54595.4049	&	14	\\
53462.4063	&	6	&	54596.6184	&	15	\\
53462.5169	&	6	&	55122.4751	&	16	\\
53464.5030	&	6	&	55122.5856	&	16	\\
53472.4471	&	6	&	55123.5786	&	16	\\
53485.4677	&	6	&	55151.3853	&	16	\\
53504.3365	&	8	&	55151.4958	&	16	\\
53504.4465	&	8	&	55151.6054	&	16	\\
53517.7983	&	9	&	55676.32502	&	16	\\
53765.5180	&	10	&	55676.43554	&	16	\\
53818.3717	&	7	&	            &      \\
\hline
\end{tabular}
\end{center}
\scriptsize {References for Table~\ref{tab:CCCom:mintimes}.
 1- Zejda (2004);
 2- Nelson (2004);
 3- Agerer \& Hubscher (2003);
4- Kim (2006);
5- Hubscher (2005);
6- Hubscher et al. (2005);
7- Hubscher (2006);
8- Pribulla et al. (2005);
9- Nelson (2006);
10- Hubscher (2007);
11- Parimucha et al. (2007);
12- Do{\u g}ru (2006);
13- Do{\u g}ru (2007);
14- Hubscher et al. (2009);
15- Dvorak (2009);
16- present study.}
\end{table}

\begin{figure}
\includegraphics[height=110mm]{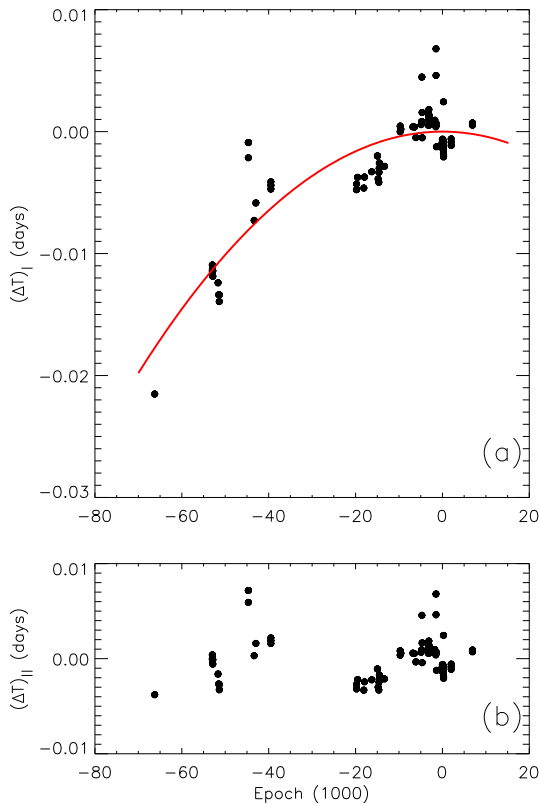} \caption{ (a) Residuals for the times of minimum
light of CC Com. The solid line is obtained with the quadratic
terms in ephemeris (Eq.~\ref{cccomEq2}). (b) The difference between the
observations and the quadratic ephemeris.}\label{Fig:CCCom:OC}
\end{figure}

The residuals shown in Fig.~\ref{Fig:CCCom:OC}a indicate a quadratic solution with a binary period
decreasing with time. In order to obtain the light elements given in
Eq.(\ref{cccomEq2}) the differential correction method is used.
By applying this equation to the times of minima given in
Table~\ref{tab:CCCom:mintimes} and by using a weighted least squares solution
we obtain
\begin{equation}\label{cccomEq2}
\begin{array}{c}
\rm{HJD\,MinI}  = 24\,54151.60676(7)  \\ + 0.22068573(9)\times E
-4.04(18)\times 10^{-12}\times E^2.
\end{array}
\end{equation}
The observed O--C values given in Fig.~\ref{Fig:CCCom:OC} are derived by using the
linear elements $T_{\rm o}$ and $P_{\rm o}$ given in Eq.
(\ref{cccomEq1}). The solid line in Fig.~\ref{Fig:CCCom:OC}a
shows a secular period  decrease of $dP/dt =-1.34\times10^{-8}$ which has been determined by using Eq.~(\ref{cccomEq2}).
The difference between our result and the results of previous studies is mainly because of the less scattered data set used in this study.
Fig.~\ref{Fig:CCCom:OC}b shows the residuals of a parabolic variation. These residuals may be assigned to a sine-like variation that can be
interpreted as a consequence of a third body or stellar magnetic activity of the components as was discussed by Yang et al. (2009).
At this point, however, we should emphasize that because of the absence of the data it is hard to confirm  definitely any sine-like variation.

\section{Simultaneous light and radial velocity curve analysis}

The light variation of CC Com has been studied by many researchers.
Rucinski (1976) solved the $\textrm{UBV}$ light curve and gave the photometric parameters of the system and indicated the
necessity of studying the system spectroscopically.
Maceroni et al. (1982) re-analyzed Rucinski's data assuming unspotted stellar model and 4500~K
temperature for the hotter component. Apart from the study of Maceroni et al. (1982),
however, almost all of the previous studies assumed the surface temperature of the hotter component to be 4300~K.
Bradstreet (1985) analyzed the $B$ and $I$ light curves and determined the absolute
elements of the system. Bradstreet applied a spotted solution and also discussed that the spotted model light
curve solution can be replaced by a model with a circumbinary gas stream.
Zhou (1988) by making use of Rucinski's $UBV$ data derived the parameters of the binary system.
In addition the author discussed the insignificance of any third light in the system. Recently, Zola et al. (2010)
solved  $\textrm{BVRI}$ light assuming fixed mass ratio by using three days observations.
Light curve variation over time has been indicated in some papers (e.g. Linnell \& Olson 1989, Quian 2001a, Yang \& Liu 2003).

We analyzed simultaneously two colour light curves of CC Com. \textsc{phoebe }(Pr{\v s}a \& Zwitter 2005),
that is based on the Wilson-Devinney-code (Wilson \& Devinney, 1971 and Wilson 1994) was used for the light curve analysis.
During the solution the light and radial velocity data points were weighted ($1/\sigma^2$) according to their individual standard errors($\sigma$).
On the other hand, during the solutions we used 613(R)+570(V) points for photometric and 60 points for RV data.
Hence, high weight assigned to the radial velocity data to avoid domination of the photometric data and to construct an equilibrium between the data.
The temperature of the primary star was adopted from Linnell \& Olson (1989), as 4300~K.
Gravity darkening coefficients and albedos are obtained from Lucy (1967) as $g_1$=$g_2$=0.32 and
from Rucinski (1969) as  $A_1$ = $A_2$ = 0.5. The logarithmic limb darkening coefficients are adopted from van Hamme (1993)
for solar composition and assumed to be equal for both stars for a given filter ($x_{1V} = x_{2V} = 0.798$, $x_{1R} = x_{2R} = 0.796$).
In addition, the radial velocities obtained by Pribulla et al. (2007) has been analysed simultaneously with the two colour photometric data.
Orbital inclination ($i$), mass ratio ($q$), temperature of the secondary component ($T_2$), separation of the components ($a$),
velocity of the center of gravity ($V_\gamma$), the monochromatic luminosity of star 1, L$_1$ and potential of the common surface ($\Omega$) were adjustable parameters.
The phase shift parameter was treated as a free parameter and almost no shift has been detected.

Some previous studies of CC Com revealed asymmetry in the maximum light.
This effect is also apparent in the light curves obtained in this study (Fig~\ref{fig1}).
In addition, similar variations, resulted from the stellar spots, are also detected during the minimum light.
Almost all of the light curves of the system presented in the literature are different from each other.
Therefore, the light-curve solutions have been done by assuming spotted models (Zola et al. 2010, Linnell \& Olson, 1989, Yakut et al. 2009).
The light curve shows asymmetry between two maxima probably because of the presence of the stellar spots.
Therefore, the presence of a spot on the primary component is assumed through analysis.
The \textsc{phoebe} code does not give accurate results for simultaneous solutions where the spot parameter is regarded as a free parameter.
In order to obtain the best spot parameters different solutions have been performed by changing the location, size, and the temperature of the spot.
Among these solutions, the one with the smallest standard deviation is regarded as a best solution for the spot parameters (Table~\ref{CCComtab2}).
Since the errors of spot parameters are not available, the errors given in Table~\ref{CCComtab2} may be smaller than the real values.
The best solution results for the spot parameters are given in Table~\ref{CCComtab2}.
On the other hand, we should emphasize the point that the spot parameters obtained from the LC solutions do not represent
a single spot at the related point but indicate the total active area on the stellar surface.
The filling factor ($f={\Omega_{\textrm{in}}-\Omega}/{\Omega_{\textrm{in}}-\Omega_{\textrm{out}}}$) from the
inner (${\Omega_{\textrm{in}}}$) to the outer critical surface (${\Omega_{\textrm{out}}}$) is estimated as 0.17.

The results derived from the light curve analysis are summarized and compared with the previous ones in Table ~\ref{CCComtab2}.
In Fig.~\ref{fig1} we show the observed data with model's estimation.

\begin{table*}
\begin{center}
\scriptsize
\caption{The photometric elements with their formal
1$\sigma$ errors for CC Com and their comparison to previous solutions of Rucinski (1976) [R76], Maceroni et al. (1982) [M82], Bradstreet (1985) [B85], Zhou (1988) [Z88], Linnell \& Olson (1989) [L89] and Zola et al. (2010) [Z10].}
\label{CCComtab2}
\begin{tabular}{llllllll}
\hline
Parameter                        &R76       &M82 ($B$; $V$)     &B85        &Z88        &L89    & Z10          & This Study       \\
\hline
Geometric parameters:            &          &                    &           &           &         &             &           \\
$i$ ${({^\circ})}$               &90        &87.92; 87.92        &90.0       &87.7       &85.2     & 84.8        & 89.8(6)   \\
$\Omega _{1} = \Omega _{2}$      &4.997     &5.045; 5.029        &5.047      &2.779      &5.168    & 2.873       & 5.017(36)  \\
$q$                              &1.919     &1.930; 1.934        &1.926      &1.70       &1.960    & 1.89        & 1.90(1)   \\
Filling factor (f \%)            &23.5      &18; 21.6            &16.7       &24         &4.4      & 18          & 17           \\
Fractional radius of primary     &0.3395    &0.3345; 0.3369      &0.3337     &0.4014     &0.3236   & 0.3294      & 0.3348(43) \\
Fractional radius of secondary   &0.4526    &0.4493; 0.4513      &0.4480     &0.4884     &0.4415   & 0.4467      & 0.4469(40) \\
Radiative parameters:            &          &                    &           &           &         &             &             \\
$T_1^*$ (K)                      &4300      &4500; 4500          &4300       &4300       &4300     & 4300        & 4300        \\
$T_2$ (K)                        &4082      &4317; 4288          &4140       &4265       &4133     & 4263        & 4200(60)    \\
Luminosity ratio:$\frac{L_1}{L_1 +L_2}$(\%) & &                 &           &           &         &             &              \\
$U$                              &          &                   &           &37         &         &             &               \\
$B$                              &          &43                 &43         &38         &         & 34          &              \\
$V$                              &          &43                 &           &38         &         & 35          & 40              \\
$R$                              &          &                   &           &           &         & 35          & 40              \\
$I$                              &          &                   &40         &           &         & 35          &                 \\
Spot on primary component\\
Colatitude   ${({^\circ})}$      &          &                   &80         &           &90       & 170.5       & 90      	 	      \\
Longitude    ${({^\circ})}$      &          &                   &           &           &-90      & 130.7       & 90		   	      \\
Spot radius  ${({^\circ})}$      &          &                   &           &           &12.5     & 50.6        & 20		         \\
Spot temperature ($T_{spot}/T_{star}$)&     &                   &0.93       &           &0.84(I)  & 0.737       & 0.92	             \\
\hline
\end{tabular}
\end{center}
\end{table*}

\section{Summary and Conclusion}
\begin{table}
\scriptsize
\begin{center}
\caption{Absolute parameters of CC Com. The standard errors
1$\sigma$ in the last digit are given in
parentheses. HC denotes the hot component and CC stand for the cool component.}\label{Tab:physpar:CCCom}
\begin{tabular}{llll}
\hline
Parameter                                        &Unit                      & CC          & HC   \\
\hline
Mass (\textbf{M})                                         &$\rm{M_{\odot}}$      & $0.717(14)$         & $0.377(8)$      \\
Radius (\textbf{R})                                       &$\rm{R_{\odot}}$      & $0.708(12)$         & $0.530(10)$      \\
Temperature (\textbf{T$_{\rm eff}$})                      &$\rm{K}$              & $4300$              & $4200(180)$    \\
Luminosity (\textbf{L})                                   &$\rm{L_{\odot}} $     & $0.138(12)$         & $0.085(7)$      \\
Surface gravity $(\log\,\textbf{g})$                      &cgs                   & $4.59$              & $4.57$          \\
Absolute bol. mag. (\textbf{M$_b$})            &mag                   & 6.90(14)            & 7.43(18)            \\
Absolute vis. mag. (\textbf{M$_V$})                &mag                   & 7.86(16)            & 8.25(20)           \\
Period change rate ($\dot{\textbf{P}}$)                   &d/yr                  &\,\,\,\,\,\,\,\,\,\,\,\,\,\,$-1.34\times10^{-8}$ &      \\
Mass transfer ratio ($\dot{\textbf{M}}$)                  &M$_\odot$/yr          &\,\,\,\,\,\,\,\,\,\,\,\,\,\,$-1.6\times10^{-8}$ &      \\
Distance (\textbf{d})                                     &pc                    &\,\,\,\,\,\,\,\,\,\,\,\,\,\,\,\,\,\,\,\,\,\,64(4)         &      \\
\hline
\end{tabular}
\end{center}
\end{table}
Period variation and light-curve analysis of the low temperature contact binary (LTCB) system CC Com have been studied.
The physical parameters of the system have been determined with the simultaneous solution of our two
colour light curves (\emph{VR}) and the spectroscopic study of Pribulla et al. (2007).

The measured physical parameters of CC Com have been presented in Table~\ref{Tab:physpar:CCCom} with their errors.
It seems that there is a good agreement between the derived physical parameters of CC Com and other LTCBs (Yakut \& Eggleton, 2005).
We estimate the distance to CC Com by using the results obtained from radial velocity and light curve analysis.
For that purpose the total brightness (V=11\fm30) and light ratio of the components have been used.
The results indicate a distance of 65 pc and 63 pc for the hot and cool components, respectively.
The mean of these estimates gives the distance of the system as 64(4) pc. This value is 20\% smaller than the value given by SIMBAD database.

We have collected and analyzed the times of minima for CC Com. The (O--C) diagram shows a downward parabola.
This property can be explained as a mass transfer from the more massive component to the less massive one.
The quadratic term of Eq.~\ref{cccomEq2} shows that the orbital period of the system decreases at a
rate of ${dP}/{dt}=1.34\times 10^{-8}$\,days/yr as a result of mass transfer rate of $1.6\times 10^{-8}$ M$_\odot$/yr.
In this study, the residuals of the period variation do not show any reliable sine-like variation (Fig.~\ref{Fig:CCCom:OC}b).
Assuming the existence of the third body we solved the data, however, because of the poor data it is hard to find any evidence for a third body in the system.
On the other hand, Yang et al. (2009) assumed a third body in the system; solved the residuals and determined the third orbit parameters.

Most of the contact binaries show the O'Connell effect in their light curves. This effect is due to large cool starspots (see for details Kalomeni et al. 2007).
The variation seen in Fig.~\ref{Fig:CCCom:OC}b can also be explained by stellar activity.
Similar residuals have been observed in many contact binary systems (e.g., XY Leo, Yakut et al. 2003).
The light curve solution indicates that 6\% of the primary star's surface is covered by a cold spot.
In this case, Applegate mechanism (Applegate 1992) can be responsible from the variation in the orbital period and the occurrence of the non-periodic change.

Different scenarios have been proposed for the evolution and structure of contact binaries.
The prevailing theory among them is the thermal relaxation oscillation (TRO) theory, proposed by Lucy (1976), Flannery (1976),  Robertson \& Eggleton (1977),
Yakut \& Eggleton (2005). The TRO model can explain successfully  the evolution, structure and the
observed properties of contact binary systems (Wang, 1994, Qian 2001b, Van Hamme, 2001, Webbing 2003,
Paczy{\'n}ski et al. 2006, Zhu et al. 2010).  Stepien (2006), also proposed a scenario explaining the
evolution of contact binary systems. In short period binaries like CC Com, angular momentum loss plays a crucial role in their evolution.
For a close detached binary system with initial parameters 1.19 $\rm{M_{\odot}}$ + 0.94 $\rm{M_{\odot}}$ and  0.75 days,
the mass of the system is estimated to decrease by $\sim 15\%$  (0.97 $\rm{M_{\odot}}$+0.83 $\rm{M_{\odot}}$)
until \textsc{RLOF} at P = 0.31 days, and it reaches the contact phase at 0.88 $\rm{M_{\odot}}$ + 0.91 $\rm{M_{\odot}}$, 0.28 days.
This scenario can explain the evolutionary stages of CC Com with TRO mode like the other LTCBs that are discussed in detail in Yakut \& Eggleton (2005).

Additionally, the evolution of short period binaries is also important in measuring the gravitational waves.
Binary systems with very short period can form gravitational waves in detectable range (Ju et al., 2000, K{\" o}se \& Yakut, 2011).
We have estimated the amplitude of the gravitational wave in CC Com binary system as ($\log (\textrm{h})$) -20.6.
Hence, the system CC Com is an important source for interferometers and its amplitude is within the detection limit of
detectors such as LISA that can detect accurate gravitational waves. The relatively close distance of the system makes it also important target to study.

\acknowledgements
This study was supported by the Turkish Scientific
and Research Council (T\"UB\.ITAK 109T047), the T\"UB\.ITAK National Observatory (TUG), and the Ege University Research Fund.
KY+VK acknowledges support by the Turkish Academy of Sciences (T{\"U}BA).
The authors thank to the anonymous referee for valuable suggestions and E. R. Pek\"unl\"u for his comments.


\begin{thebibliography}{}

\bibitem{} Agerer, F., Hubscher, J.: 2003, IBVS 5484, 1
\bibitem{} Applegate, J.H.: 1992, ApJ 385, 621
\bibitem{} Bradstreet, D.H.: 1985, Astrophys. J. Suppl. 58, 413
\bibitem{} Breinhorst, R.A., Hoffmann, M.: 1982, Ap\&SS 86, 107
\bibitem{} Do{\u g}ru, S.S., Do{\u g}ru, D., Erdem, A. et al.: 2006, IBVS 5707, 1
\bibitem{} Do{\u g}ru, S.S., Do{\u g}ru, D., D{\" o}nmez, A.: 2007, IBVS 5795, 1
\bibitem{} Dvorak, S.W.: 2009, IBVS 5870, 1
\bibitem{} Flannery, B.P.: 1976, ApJ 205, 217
\bibitem{} Hoffmeister, C.: 1964, AN 288, 49
\bibitem{} Hubscher, J.: 2005, IBVS 5643, 1
\bibitem{} Hubscher, J.: Paschke A., Walter F., 2005, IBVS 5657, 1
\bibitem{} Hubscher, J.: Paschke A., Walter F., 2006, IBVS 5731, 1
\bibitem{} Hubscher, J.: 2007, IBVS 5802, 1
\bibitem{} Hubscher, J., Steinbach, H.M., Walter, F.: 2009, IBVS 5874,1
\bibitem{} Ju, L., Blair, D.~G., Zhao, C.: 2000, RPPh 63, 1317
\bibitem{} Kalomeni, B., Yakut, K., Keskin, et al.: 2007, AJ 134, 642
\bibitem{} Kim, C.H., Lee, C.U., Yoon, Y.N., et al.: 2006, IBVS 5694, 1
\bibitem{} K{\"o}se, O., \& Yakut, K.: 2011, in preparation
\bibitem{} K{\"o}se, O., Keskin, V., Yakut, K.: 2009, Ap\&SS 323, 75
\bibitem{} Linnell, A.P., Olson, E.C.: 1989, ApJ 343, 909
\bibitem{} Lucy, L.B.: 1967, Zeit. Astrophys. 65, 89
\bibitem{} Lucy, L.B.: 1976, ApJ 205, 208
\bibitem{} Maceroni, C., Milano, L., Russo, G.: 1982, A\&AS 49, 123
\bibitem{} McLean, B.J., Hilditch, R.W.: 1983, MNRAS 203, 1
\bibitem{} Nelson, R.H.: 2004, IBVS 5493, 1
\bibitem{} Nelson, R.H.: 2006, IBVS 5672, 1
\bibitem{} Paczy{\'n}ski, B., Szczygie{\l}, D.~M., Pilecki, B., Pojma{\'n}ski, G., 2006, MNRAS 368, 1311
\bibitem{} Parimucha, S., et al.: 2007, IBVS 5777, 1
\bibitem{} Pribulla, T., et al.: 2005, IBVS 5668, 1
\bibitem{} Pribulla, T., Rucinski, S.M., Conidis, G., et al.: 2007, AJ 133, 1977
\bibitem{} Pr{\v s}a, A., \& Zwitter, T.: 2005, ApJ 628, 426
\bibitem{} Robertson, J.A., Eggleton P.P.: 1977, MNRAS 179, 359
\bibitem{} Rucinski, S.M.: 1969, Acta Astron. 19, 245
\bibitem{} Rucinski, S.M.: 1976, PASP 88, 777
\bibitem{} Rucinski, S.M., Whelan J.~A.~J., Worden S.~P.: 1977, PASP 89, 684
\bibitem{} Stepien, K.: 2006, AcA 56, 199
\bibitem{} Qian, S.: 2001a, Ap\&SS 278, 415
\bibitem{} Qian, S.: 2001b, MNRAS 328, 914
\bibitem{} van Hamme, W.: 1993, AJ 106, 2096
\bibitem{} van Hamme, W., Samec, R.G., Gothard, N.W. et al.: 2001, AJ 122, 3436
\bibitem{} Wang, J.M.: 1994, ApJ 434, 277
\bibitem{} Webbink, R.F.: 2003, ASPC 293, 76
\bibitem{} Wilson, R.E. \& Devinney, E. J.: 1971, ApJ 166, 605
\bibitem{} Wilson, R.E.: 1994, PASP 106, 921
\bibitem{} Yakut, K., \& Eggleton, P.~P.: 2005, ApJ 629, 1055
\bibitem{} Yakut, K., et al.: 2009, A\&A 503, 165
\bibitem{} Yakut, K., {\.I}bano{\u g}lu, C., Kalomeni, B., et al.: 2003, A\&A 401, 1095
\bibitem{} Yang, Y., Liu, Q.: 2003, PASP 115, 748
\bibitem{} Yang, Y.G., L{\"u}, G.L., Yin, et al.: 2009, AJ 137, 236
\bibitem{} Zejda, M.: 2004, IBVS 5583, 1
\bibitem{} Zhou, H.N.: 1988, Ap\&SS 141, 199
\bibitem{} Zhu, L., Qian, S.B., Mikul{\'a}{\v s}ek, Z., et al.: 2010, AJ 140, 215
\bibitem{} Zola, S., Gazeas, K., Kreiner, J.M., et al.:  2010, MNRAS 408, 464
\end{thebibliography}
\end{document}